\documentclass[11pt,a4paper]{article}
\usepackage{jcappub,natbib}
\usepackage{braket}
\input{colordvi.tex}
\title{Parity violation in the CMB bispectrum by a rolling pseudoscalar}
\author[a,b]{Maresuke Shiraishi,}
\author[a,b]{Angelo Ricciardone}
\author[c]{and Shohei Saga}
\affiliation[a]{Dipartimento di Fisica e Astronomia ``G. Galilei'', Universit\`a degli Studi di Padova, via Marzolo 8, I-35131, Padova, Italy}
\affiliation[b]{INFN, Sezione di Padova, via Marzolo 8, I-35131, Padova, Italy}
\affiliation[c]{Department of Physics and Astrophysics, Nagoya University, 
Nagoya, Aichi, 464-8602, Japan}

\abstract{%
We investigate parity-violating signatures of temperature and polarization bispectra of the cosmic microwave background (CMB) in an inflationary model where a rolling pseudoscalar produces large equilateral tensor non-Gaussianity. By a concrete computation based on full-sky formalism, it is shown that resultant CMB bispectra have nonzero signals in both parity-even ($\ell_1 + \ell_2 + \ell_3 = {\rm even}$) and parity-odd ($\ell_1 + \ell_2 + \ell_3 = {\rm odd}$) spaces, and are almost uncorrelated with usual scalar-mode equilateral bispectra. These characteristic signatures and polarization information help to detect such tensor non-Gaussianity. Use of both temperature and E-mode bispectra potentially improves of 400\% the detectability with respect to an analysis with temperature bispectrum alone. Considering B-mode bispectrum, the signal-to-noise ratio may be able to increase by 3 orders of magnitude. We present the $1\sigma$ uncertainties of a parameter depending on a coupling constant and a rolling condition for the pseudoscalar expected in the {\it Planck} and the proposed PRISM experiments.
}

\begin{document}

\maketitle
\flushbottom

\section{Introduction}

Cosmological parity violation may be a key indicator of UV theories of gravity and early Universe models, and has been well-studied from both theoretical and observational sides (e.g., refs.~\cite{Lue:1998mq, Alexander:2004wk, Lyth:2005jf, Saito:2007kt, Satoh:2007gn, Takahashi:2009wc, Gluscevic:2010vv, Gruppuso:2010nd, Barnaby:2010vf, Koivisto:2010fk, Groeneboom:2010fn, Sorbo:2011rz, Barnaby:2011vw, Dimopoulos:2012av, Gluscevic:2012me, Barnaby:2012xt, Grain:2012cx, Wang:2012fi, Shiraishi:2013vja, Ade:2013nlj}). Nowadays the investigation of the connections between the parity violation and tensor non-Gaussianity has attracted attention \cite{Maldacena:2011nz, Soda:2011am, Shiraishi:2011st, Shiraishi:2012sn, Zhu:2013fja, Cook:2013xea}. Such non-Gaussianity imprints new types of distinguishable signatures in temperature and polarization bispectra of the cosmic microwave background (CMB), e.g., temperature auto-bispectrum in $\ell_1 + \ell_2 + \ell_3 = {\rm odd}$ or B-mode auto-bispectrum in $\ell_1 + \ell_2 + \ell_3 = {\rm even}$ \cite{Kamionkowski:2010rb, Shiraishi:2011st, Shiraishi:2012sn}.

Recently, ref.~\cite{Barnaby:2012xt} has proposed an inflationary model where a rolling pseudoscalar, gravitationally coupled to the inflaton, amplifies the vacuum fluctuations of a $U(1)$ gauge field. This gauge field can add extra signals in power spectrum and bispectrum of curvature perturbations and of gravitational waves in addition to normal signals generated by the inflaton. The introduction of a second (pseudoscalar) field minimizes the amount of scalar perturbations and an interesting gravitational wave signal can be obtained without conflicting with the bounds on non-Gaussianity from the scalar perturbations. Resulting gravitational waves are chiral, can produce TB and EB correlations and parity-violating non-Gaussianities. Investigating these characteristic observables is a meaningful way to judge the validity of this model. More recently, ref.~\cite{Cook:2013xea} has found that in this model the non-Gaussianity of gravitational waves is ${\cal O}(10^3)$ times than the curvature non-Gaussianity. This implies the existence of sizable CMB bispectrum unlike usual scalar-mode one. The authors have evaluated the magnitude of resultant temperature auto-bispectrum through an approximation based on flat-sky formalism, and have translated a current bound on the equilateral nonlinearity parameter into a rough bound on the pseudoscalar coupling. Their analysis is a reasonable way to evaluate the signals roughly. However, their flat-sky analysis may be no longer appropriate on large scales where the tensor mode is effective. As we will show, the tensor non-Gaussianity in this model breaks parity invariance asymmetrically and creates separate signals in both parity-even ($\ell_1 + \ell_2 + \ell_3 = {\rm even}$) and parity-odd ($\ell_1 + \ell_2 + \ell_3 = {\rm odd}$) spaces. It may be hard to evaluate such signatures precisely in the flat-sky formalism which is based on non-discrete $\ell$ space. Furthermore, we expect that inclusion of the polarization bispectra can improve detectability drastically since, in this case, the tensor mode is a major source of non-Gaussianity \cite{Shiraishi:2013vha}.

In this paper, we present a concrete study of the temperature and polarization bispectra generated from a rolling pseudoscalar. Firstly, on the basis of a full-sky formalism with full radiation transfer dependence \cite{Shiraishi:2010kd}, we construct a general form for the CMB bispectra. The primordial tensor bispectrum is given by a non-separable form between three wave numbers and it makes the computation of the CMB bispectra quite difficult. We solve this by replacing it with a reconstructed separable one. Through technical treatments of $\ell$ dependence in the full-sky formulation, we confirm that resulting CMB bispectra do not vanish both in the parity-even and parity-odd $\ell$ spaces. Next, we estimate the detectability of the tensor non-Gaussianity for cases with the auto- and cross-bispectra between the temperature and E-mode anisotropies, and with the B-mode auto-bispectrum alone. In the analysis with the temperature and E-mode bispectra, we assume the existence of a contamination by the standard equilateral non-Gaussianity, while we show that it is a negligible effect. For both cases, we assume the {\it Planck} and the proposed PRISM experiments. Then, we show that considering the polarization information and both the parity-even and parity-odd signals improves the detectability. We also summarize the expected $1\sigma$ errors of a model parameter determined by a coupling constant and a rolling condition of the pseudoscalar field. 

This paper is organized as follow. In the next section, we review an inflationary model with a rolling pseudoscalar by following ref.~\cite{Barnaby:2012xt}. In section~\ref{sec:hhh}, we compute the primordial tensor bispectrum and find its reconstructed form which is useful in CMB computation. In section~\ref{sec:CMB_bis}, we produce a full-sky form for the CMB temperature and polarization bispectra and analyze their behaviors. Section~\ref{sec:Fisher} presents Fisher matrix analysis for estimating the detectability of the tensor non-Gaussianity, and the final section is devoted to summary and discussion of our results.

\section{Gauge field amplification by a rolling pseudoscalar}\label{sec:pseudo}

In this paper, we consider a model where in addition to a standard inflationary sector, we have a (hidden) sector with a rolling pseudoscalar $\chi$ coupled to a $U(1)$ gauge field $A_\mu$, whose Lagrangian is given by  
\begin{eqnarray}
 {\cal L} =  
-\frac{1}{2} (\partial \phi)^2 - V(\phi) - \frac{1}{2} (\partial \chi)^2 - U(\chi) - \frac{1}{4}F_{\mu \nu} F^{\mu \nu} - \frac{\chi}{4f} F_{\mu \nu} \tilde{F}^{\mu \nu} ~, \label{eq:action}
\end{eqnarray}
where $f$ is a coupling constant like an axion decay constant and $F_{\mu \nu} \equiv \partial_\mu A_\nu - \partial_\nu A_\mu$ is the field strength and $\tilde{F}_{\mu \nu}$ its dual \cite{Barnaby:2012xt, Cook:2013xea}. In this model, a successful slow-roll inflation occurs owing to an inflaton potential $V(\phi)$, and $\chi$ contributes to the generation of curvature and tensor perturbations through gravitational interaction with the gauge field. Such a scenario is different from the case in which a direct coupling between the inflaton and the gauge field is present \cite{Barnaby:2011vw}. In that case the coupling is much stronger than the gravitational one and scalar curvature fluctuations are sourced with much more efficiency than gravitational waves  \cite{Barnaby:2012xt}. Observed power spectra of curvature and tensor perturbations will consist of both these gauge-field modes and considerable normal modes generated in the slow-roll regime, which are expressed as ${\cal P} \equiv \frac{H^2}{8 \pi^2 \epsilon M_P^2}$ and ${\cal P}_h = 16 \epsilon {\cal P}$ with $H$, $\epsilon$ and $M_P \equiv 1/\sqrt{8 \pi G}$ being the Hubble parameter, the slow-roll parameter for the inflaton and the reduced Planck mass, respectively. On the other hand, the gauge-field contributions can dominate over the bispectrum signals owing to the slow-roll suppression of the normal-mode non-Gaussianities \cite{Barnaby:2012xt, Cook:2013xea}. This implies that we can obtain tight constraints on this model from CMB bispectrum analysis.  

We shall analyze the dynamics of the gauge field in the Coulomb gauge $A_0 = 0$ and $\nabla \cdot {\bf A} = 0$. Then, equation of motion of the gauge field reads 
\begin{eqnarray}
{\bf A}'' - \nabla^2 {\bf A} - \frac{\chi'}{f} \nabla \times {\bf A} = 0 ~,
\end{eqnarray}
where $~'~ \equiv \partial/ \partial \tau$ denotes conformal time derivative. To solve this, we move to a quantization process in Fourier space, reading 
\begin{eqnarray}
A_i(\tau, {\bf x}) &=& 
\int \frac{d^3{\bf k}}{(2\pi)^{3/2}}
\sum_{\lambda = \pm 1} 
v_{\lambda}(\tau, {\bf k}) 
\epsilon_i^{(\lambda)}({\bf k})
 e^{i {\bf k} \cdot {\bf x}} ~, \\
v_{\lambda}(\tau, {\bf k}) 
&=& a_{\lambda}({\bf k}) A_\lambda (\tau, {\bf k}) 
+ a_{\lambda}^\dagger(- {\bf k}) A_\lambda^* (\tau, -{\bf k}) ~,
\end{eqnarray} 
where creation and annihilation operators satisfy 
\begin{eqnarray}
\left[a_\lambda({\bf k}), a_{\lambda'}^\dagger({\bf k'})\right] = \delta_{\lambda \lambda'}\delta^{(3)}({\bf k} - {\bf k'})~,
\end{eqnarray}
and $\epsilon_i^{(\pm 1)}$ is a divergenceless polarization vector (for details see appendix~\ref{appen:polarization}).\footnote{
We use the Fourier transform convention as 
\begin{eqnarray}
f({\bf x}) = \int \frac{d^3 {\bf k}}{(2\pi)^{3/2}} 
f({\bf k}) e^{i {\bf k} \cdot {\bf x}} ~.
\end{eqnarray}
The polarization vector $\epsilon_i^{(\pm 1)}(\hat{\bf k})$ is equivalent to $\epsilon_{\pm}^i({\bf k})$ in refs.~\cite{Barnaby:2011vw, Sorbo:2011rz, Barnaby:2012xt, Cook:2013xea}.
} 
Then assuming rolling condition like $\dot{\chi} \simeq {\rm const.}$ leads to an explicit form of $A_+$:
\begin{eqnarray}
A_+(\tau, k) \simeq \frac{1}{\sqrt{2k}} 
\left( -\frac{k\tau}{2\xi} \right)^{1/4} 
e^{\pi \xi - 2 \sqrt{-2 \xi k \tau}} ~,
\end{eqnarray}
where $\xi \equiv \frac{\dot{\chi}}{2fH}$ with $~\dot{}~ \equiv d/dt$ being physical time derivative, and $\xi > 0$ is assumed without loss of generality. We are interested in the situation where the gauge field may give observable effects on cosmological perturbations, namely $\xi \gtrsim {\cal O}(1)$. Then, this solution will perform well for all interesting scales. Note that $A_+$ is exponentially amplified as $\xi$ becomes large, due to tachyonic instability, while $A_-$ has no amplification mechanism and is negligible in comparison to $A_+$ \cite{Barnaby:2011vw}. Owing to this chiral property, the tensor non-Gaussianity and resultant CMB bispectra break parity invariance and the signal can be distinguishable from the vacuum one since one gravity wave helicity is produced in a much stronger way than the other.

\section{Parity-violating tensor non-Gaussianity}\label{sec:hhh}

The produced gauge field quanta give rise to scalar and tensor modes, giving rise to non-standard power spectra and bispectra. Very interestingly, due to helicity conservation, the tensor non-Gaussianity has a larger amplitude in comparison with the scalar one that can be considered negligible \cite{Cook:2013xea}. In this section, we shall formulate such tensor non-Gaussianity to estimate its CMB signals in our convention. Then, we will confirm the consistency of our results with ref.~\cite{Cook:2013xea}. 

\subsection{Primordial tensor bispectrum} 

The tensor metric perturbation, which is defined in $\delta g_{ij}^{(T)} = a^2 h_{ij}$ with $a$ being the scale factor, obeys the Einstein equation: 
\begin{eqnarray}
h_{ij}'' + 2 \frac{a'}{a} h_{ij}' - \nabla^2 h_{ij} 
= - \frac{2a^2}{M_P} (E_i E_j + B_i B_j)^{TT} ~,
\end{eqnarray}
where ${\bf E} = - {\bf A}' / a^2$ and ${\bf B} = \nabla \times {\bf A} / a^2$ are electric and magnetic parts of the gauge field, and $TT$ denotes transverse and traceless elements. Here, the source term arises not from the $F\tilde{F}$ term but from the $FF$ term in eq.~(\ref{eq:action}). Parity-violating information of the gauge field is transmitted to the tensor metric perturbation through this source term. In Fourier space, a solution is given by the Green function $G_k$ as \cite{Cook:2013xea}  
\begin{eqnarray}
h_{ij}(\tau, {\bf x}) &=& 
\int \frac{d^3{\bf k}}{(2\pi)^{3/2}}
\sum_{\lambda = \pm 2} 
h_{\bf k}^{(\lambda)}(\tau) 
e_{ij}^{(\lambda)}(\hat{\bf k})
 e^{i {\bf k} \cdot {\bf x}} ~, \\
h_{\bf k}^{(\lambda)}(\tau) 
&=& -\frac{2H^2}{M_P^2} \int d \tau' G_k (\tau, \tau') \tau'^2
\int \frac{d^3 {\bf k'}}{(2\pi)^{3/2}} d^3 {\bf k''}  \frac{1}{2} e_{ij}^{(-\lambda)}(\hat{\bf k}) 
\nonumber \\ 
&&\times 
\left[ {\cal E}_{i}(\tau', {\bf k'}) {\cal E}_j(\tau', {\bf k''}) 
+ {\cal B}_i(\tau', {\bf k'}) {\cal B}_j(\tau', {\bf k''}) \right] \delta({\bf k'} + {\bf k''} - {\bf k}) ~,
\end{eqnarray}
where $e_{ij}^{(\pm 2)}$ is the transverse and traceless polarization tensor defined in appendix~\ref{appen:polarization}\footnote{Our polarization tensor $e_{ij}^{(\pm 2)}(\hat{\bf k})$ is equal to $2 \Pi_{\mp}^{ij}({\bf k})$ in refs.~\cite{Sorbo:2011rz, Cook:2013xea}.} and 
\begin{eqnarray}
G_k(\tau, \tau') = 
\frac{1}{k^3 \tau'^2 }
\left[ (1 + k^2 \tau \tau') \sin k(\tau-\tau') 
+ k(\tau' - \tau) \cos k(\tau-\tau')
\right] \Theta(\tau- \tau') ~. 
\end{eqnarray}
${\cal E}_i$ and ${\cal B}_i$ represent the dependence on $E_i$ and $B_i$ in Fourier space, namely
\begin{eqnarray}
{\cal E}_i(\tau, {\bf k}) &\equiv& 
v'_{+}(\tau, {\bf k}) \epsilon_i^{(+)}({\bf k}) ~, \\
{\cal B}_i(\tau, {\bf k}) &\equiv& 
k v_{+}(\tau, {\bf k}) \epsilon_i^{(+)}({\bf k})  ~,
\end{eqnarray}
where we ignore $A_-$ for its smallness. Then, the tensor bispectrum is formed by 6-point functions of ${\cal E}_i$ and ${\cal B}_i$. Computing it on superhorizon scales ($-k\tau \to 0$) using the Wick's theorem and conventions of the polarization vector and tensor (appendix~\ref{appen:polarization}), yields 
\begin{eqnarray}
\Braket{\prod_{n=1}^3 h^{(\lambda_n)}_{\bf k_n}} 
&=& \left[ \prod_{n=1}^3 \int \frac{d^3{\bf k_n'}}{(2\pi)^{3/2}} \right]
\delta({\bf k_1} - {\bf k_1'} + {\bf k_3'}) 
\delta({\bf k_2} - {\bf k_2'} + {\bf k_1'})
\delta({\bf k_3} - {\bf k_3'} + {\bf k_2'})
\nonumber \\
&&\times 
\epsilon_a^{(+)}(\hat{\bf k_1'}) \epsilon_d^{(-)}(\hat{\bf k_1'}) 
\epsilon_e^{(+)}(\hat{\bf k_3'}) \epsilon_b^{(-)}(\hat{\bf k_3'})
\epsilon_c^{(+)}(\hat{\bf k_2'}) \epsilon_f^{(-)}(\hat{\bf k_2'}) \nonumber \\
&&\times \frac{1}{2} e_{ab}^{(-\lambda_1)}(\hat{\bf k_1}) 
 \frac{1}{2} e_{cd}^{(-\lambda_2)}(\hat{\bf k_2}) 
 \frac{1}{2} e_{ef}^{(-\lambda_3)}(\hat{\bf k_3}) 
{\cal F}(k_1', k_2', k_3') \nonumber \\ 
&\equiv& (2\pi)^{-3/2}B^{\lambda_1 \lambda_2 \lambda_3}_{{\bf k_1} {\bf k_2} {\bf k_3}} 
\delta^{(3)}\left( \sum_{n=1}^3 {\bf k_n} \right)
~, \label{eq:hhh_exact}
\end{eqnarray}
where 
\begin{eqnarray}
{\cal F}(k_1', k_2', k_3') 
&\equiv& \left(- \frac{2H^2}{M_P^2} \right)^3 
\left[ \prod_{n=1}^3\lim_{-k_n \tau \to 0} \int_{-\infty}^0 d\tau_n 
\tau_n^2 G_{k_n}(\tau, \tau_n)
\right] 
\nonumber \\  
&&\times  
{\cal A}_h(\tau_1, k_1', k_3') 
{\cal A}_h(\tau_2, k_2', k_1')  
{\cal A}_h(\tau_3, k_3', k_2') ~, \label{eq:calF} 
\end{eqnarray}
and ${\cal A}_h$ is given by
\begin{eqnarray}
{\cal A}_{h}(\tau, k, q) 
&\equiv& 2 \left[ A_+'(\tau, k) 
A_+'(\tau, q) 
+ kq A_+(\tau, k) A_+(\tau, q) \right] ~. \label{eq:calAh}
\end{eqnarray}
We are interested in the bispectrum signatures for $\xi \gtrsim {\cal O}(1)$. In this condition, eq.~(\ref{eq:calAh}) becomes 
\begin{eqnarray}
{\cal A}_{h}(\tau, k, q) 
\simeq \left( \sqrt{-\frac{kq\tau}{2\xi}} - \sqrt{- \frac{2\xi}{\tau}} \right) 
[kq]^{1/4}e^{2\pi \xi}e^{-2\sqrt{-2\xi \tau}(\sqrt{k}+\sqrt{q})}~,
\end{eqnarray}
and hence $\cal F$ is analytically evaluated as  \cite{Barnaby:2011vw}
\begin{eqnarray}
{\cal F}(k_1', k_2', k_3')
\simeq
\frac{\Gamma(7)^3}{3^3 2^{24}} 
\frac{H^6}{M_P^6} 
 \frac{e^{6\pi \xi}}{\xi^9}
 \frac{(k_1' k_2' k_3')^{1/2}}
{[(\sqrt{k_1'} + \sqrt{k_2'}) (\sqrt{k_2'} + \sqrt{k_3'}) 
(\sqrt{k_3'} + \sqrt{k_1'})]^7} ~.
\end{eqnarray}
This form is equivalent to that found in \cite{Cook:2013xea}. From numerical evaluation, we confirm that this tensor bispectrum resembles closely the usual equilateral shape, and the bispectrum amplitude in the equilateral configuration ($k_1 = k_2 = k_3$) can be expressed as 
\begin{eqnarray}
\Braket{\prod_{n=1}^3 h_{\bf k_n}^{(+2)}}_{k_n \to k} \simeq 6 \times 10^{-10}
\frac{H^6}{M_P^6} \frac{e^{6\pi\xi}}{\xi^9}
\frac{\delta({\bf k_1} + {\bf k_2} + {\bf k_3})}{k^6}~.
\end{eqnarray}
This is in agreement with the result in ref.~\cite{Cook:2013xea}. We also confirm the smallness of the other spin modes, reading $B^{+2+2+2} \sim 10^2 B^{+2+2-2} \sim 10^5 B^{+2-2-2} \sim 10^5 B^{-2-2-2}$, and hence we shall ignore these contributions in the rest of the paper.

\subsection{Reconstruction for CMB bispectrum}

An exact form of the tensor bispectrum (\ref{eq:hhh_exact}) has three convolutions with respect to ${\bf k_n'}$ due to the 6-point functions of ${\cal E}_i$ and ${\cal B}_i$. This implies that resultant CMB bispectrum has also three convolutions in $\ell$ space, which corresponds to the 1-loop computation \cite{Shiraishi:2012rm, Shiraishi:2012sn}. The numerical computation of such CMB bispectrum is quite hard due to the non-separability of the $k$ integrals. Accordingly, we introduce an approximate separable form without convolutions, which is reconstructed from the exact bispectrum.

The radial function ${\cal F}$ has three poles, i.e., $k_1' = k_2' = 0$, $k_2' = k_3' = 0$ and $k_3' = k_1' = 0$, and contributions around these poles may dominate over total signals. Evaluating these contributions in the similar way as refs.~\cite{Shiraishi:2012rm, Shiraishi:2012sn} yields
\begin{eqnarray}
B^{+2 +2 +2}_{{\bf k_1} {\bf k_2} {\bf k_3}}
&\approx& 
(2\pi)^{-3}
 \frac{\Gamma(7)^{3}}{3^{3}2^{27}}
 \frac{H^6}{M_P^6} \frac{e^{6\pi \xi}}{\xi^9} 
\delta^{(3)}\left(\sum_{n=1}^3 {\bf k_n}\right) 
\frac{4\pi}{3} 
e_{ab}^{(-2)}(\hat{\bf k_1}) 
e_{bc}^{(-2)}(\hat{\bf k_2}) 
 e_{ca}^{(-2)}(\hat{\bf k_3}) 
\nonumber \\ 
&&\times \frac{\ln\left( \frac{k_{\rm max}}{k_{\rm min}} \right)}{2^7}
\left[
\frac{ \sqrt{k_{1}k_{2}} }
{(\sqrt{k_1}+\sqrt{k_2})^{14}} 
 + \frac{ \sqrt{k_2 k_3} }
{(\sqrt{k_2}+\sqrt{k_3})^{14}}
+ \frac{ \sqrt{k_3 k_1}}
{(\sqrt{k_3}+\sqrt{k_1})^{14}}
\right]~. 
\end{eqnarray}
where we have used relationships between the polarization vector and tensor (appendix~\ref{appen:polarization}) and 
\begin{eqnarray}
 \int \frac{k_1'^2 dk_1'}{(2\pi)^{9/2}} {\cal F}(k_1' \sim k_2' \ll k_3')  
&=& ( 2\pi)^{-9/2}  \frac{\Gamma(7)^{3}}{3^{3}2^{24}}
 \frac{H^6}{M_P^6}
\frac{e^{6\pi\xi}}{\xi^{9}}
 \frac{\ln\left( \frac{k_{\rm max}}{k_{\rm min}} \right)}{2^7}
\frac{\sqrt{k'_{2}k'_{3}}}{(\sqrt{k'_{2}}+\sqrt{k'_{3}})^{14}} ~, \\
\int d^2\hat{\bf k'} \epsilon_a^{(1)}(\hat{\bf k'}) \epsilon_b^{(-1)}(\hat{\bf k'}) &=&
\frac{4\pi}{3} \delta_{ab} ~.
\end{eqnarray}
This form produces an almost exact spin and angle dependence. However, since it is not separable with respect to $k_n$, we have to alter it to a separable form. Then, we shall replace the non-separable part with the usual equilateral template as 
\begin{eqnarray}
B^{+2 +2 +2}_{{\bf k_1} {\bf k_2} {\bf k_3}} 
&\approx& N {\cal P}^3 X^3 B^{\rm eq}_{k_1 k_2 k_3} 
e_{ab}^{(-2)}(\hat{\bf k_1})
e_{bc}^{(-2)}(\hat{\bf k_2})
e_{ca}^{(-2)}(\hat{\bf k_3}) ~, \label{eq:hhh_approx} \\ 
B^{\rm eq}_{k_1 k_2 k_3} 
&=& - \frac{1}{k_1^3 k_2^3} - \frac{1}{k_2^3 k_3^3} - \frac{1}{k_3^3 k_1^3} - \frac{2}{k_1^2 k_2^2 k_3^2} 
\nonumber \\
&&
+ \frac{1}{k_1 k_2^2 k_3^3} + \frac{1}{k_1 k_3^2 k_2^3}
+ \frac{1}{k_2 k_3^2 k_1^3} + \frac{1}{k_2 k_1^2 k_3^3}
+ \frac{1}{k_3 k_1^2 k_2^3} + \frac{1}{k_3 k_2^2 k_1^3}
~. \label{eq:bis_eq}
\end{eqnarray}
where ${\cal P}=\frac{H^2}{8\pi^{2}\epsilon M_{P}^2}$ and 
\begin{eqnarray}
X \equiv \epsilon \frac{e^{2\pi\xi}}{\xi^3}~. \label{eq:X_def}
\end{eqnarray} 

To check the shape resemblance between exact (\ref{eq:hhh_exact}) and reconstructed (\ref{eq:hhh_approx}) bispectra, we introduce the shape correlator defined as 
\begin{eqnarray}
r = \frac{B^{\rm ex} \cdot B^{\rm rec}}
{ \sqrt{(B^{\rm ex} \cdot B^{\rm ex}) (B^{\rm rec} \cdot B^{\rm rec})} } ~,
\end{eqnarray}
where a scalar product is defined as
\begin{eqnarray}
B \cdot B' &\equiv& \int_0^1 dx_2 \int_{1-x_2}^1 dx_3 (x_2 x_3)^4  B_{1, x_2, x_3} B_{1, x_2, x_3}' ~.
\end{eqnarray}
A numerical evaluation yields $r = 0.98$, which guarantees a consistency between the bispectra. Then, the normalization factor can be estimated as   
\begin{eqnarray}
N = \frac{B^{\rm ex} \cdot B^{\rm rec}(N = 1)}{B^{\rm rec}(N=1) \cdot B^{\rm rec}(N = 1)} = 4.3174 \times 10^{-3} ~.
\end{eqnarray}

In the next section, we adopt this reconstructed bispectrum in the computation of the CMB bispectra.

\section{CMB temperature and polarization bispectra} \label{sec:CMB_bis}

In this section, let us analyze CMB signatures of the parity-violating tensor non-Gaussianity discussed in the previous section. 

Tensor metric perturbations, which are generated during inflation and stretched beyond horizon, re-enter horizon around recombination and create both temperature and polarization anisotropies. Major signals in the temperature anisotropy appear on large scales ($\ell \lesssim 100$) due to the Integrated Sachs Wolfe (ISW) effect after recombination. On the other hand, the polarization anisotropies are generated through Thomson scattering at both recombination and reionization, and have peaks at corresponding scales, namely $\ell\sim 100$ and $10$, respectively \cite{Pritchard:2004qp}.

In general, the CMB anisotropies are quantified through a multipole expansion, i.e., $\frac{\Delta {\cal X}(\hat{\bf n})}{\cal X} = \sum_{\ell m} a_{\ell m}^{\cal X} Y_{\ell m}(\hat{\bf n})$. Here the superscript ${\cal X}$ denotes the temperature ($I$), E-mode ($E$) and B-mode ($B$) fields. Then, each multipole coefficient can be expressed as \cite{Shiraishi:2010kd, Shiraishi:2010sm}
\begin{eqnarray}
a_{\ell m}^{\cal X} &=& 4\pi (-i)^\ell \int_0^\infty \frac{k^2 dk}{(2\pi)^{3/2}}
 {\cal T}_\ell^{\cal X}(k) \sum_{\lambda = {\pm 2}} 
\left(\frac{\lambda}{2}\right)^x 
\int d^2\hat{\bf k}
h^{(\lambda)}_{\bf k} {}_{-\lambda}Y^*_{\ell m}(\hat{\bf k})~, \label{eq:h_lm} 
\end{eqnarray}
where $x$ discriminates parities of three modes: $x = 0$ for ${\cal X} = I,E$
and $x=1$ for ${\cal X} = B$, and ${\cal T}_{\ell}^{\cal X}$ is a radiation transfer function yielding the $\ell$ dependence as mentioned above. Accordingly, a formula for the CMB bispectra induced by the tensor non-Gaussianity reads 
\begin{eqnarray}
\Braket{\prod_{n=1}^3 a_{\ell_n m_n}^{{\cal X}_n}} 
&=& \left[\prod_{n=1}^3 4\pi (-i)^{\ell_n} \int_0^\infty \frac{k_n^2 dk_n}{(2\pi)^{3/2}}
 {\cal T}_{\ell_n}^{{\cal X}_n}(k_n) \right] \nonumber \\ 
&&\times (2\pi)^{-3/2} 
\left[ \prod_{n=1}^3 \int d^2\hat{\bf k_n} {}_{-2}Y^*_{\ell_n m_n}(\hat{\bf k_n}) \right]
B^{+2 +2 +2}_{{\bf k_1} {\bf k_2} {\bf k_3}} 
\delta^{(3)}\left( \sum_{n=1}^3 {\bf k_n} \right)
 ~. \label{eq:cmb_bis_form}
\end{eqnarray}
Here we note that dependence on $x$ disappears due to the $\lambda_n = +2$ polarizing nature of the tensor bispectrum. This equation involves integrals of angle-dependent parts in the tensor bispectrum (\ref{eq:hhh_approx}) and the delta function as shown in appendix~\ref{appen:polarization}. Dealing with these angular integrals by using Wigner symbols as in ref.~\cite{Shiraishi:2010kd}, yields a reduced form  
\begin{eqnarray}
\Braket{\prod_{n=1}^3 a_{ \ell_n m_n}^{{\cal X}_n}} 
&=& B_{\ell_1 \ell_2 \ell_3}^{{\cal X}_1 {\cal X}_2 {\cal X}_3} \left(
  \begin{array}{ccc}
  \ell_1 & \ell_2 & \ell_3 \\
   m_1 & m_2 & m_3
  \end{array}
 \right) ~, \\
B_{\ell_1 \ell_2 \ell_3}^{{\cal X}_1 {\cal X}_2 {\cal X}_3} &=& 
- \frac{( 8\pi )^{3/2} }{10} \sqrt{\frac{7}{3}} 
N {\cal P}^3 X^3  
\left[ \prod_{n=1}^3 \sum_{L_n} (-1)^{\frac{L_n}{2}} (-i)^{\ell_n} I_{\ell_n L_n 2}^{2 0 -2} \right]
I_{L_1 L_2 L_3}^{0 \ 0 \ 0}
\left\{
  \begin{array}{ccc}
  \ell_1 & \ell_2 & \ell_3 \\
   L_1 & L_2 & L_3 \\
   2 & 2 & 2 \\
  \end{array}
 \right\}
\nonumber \\ 
&&\times 
\int_0^\infty r^2 dr 
\left[\prod_{n=1}^3 \frac{2}{\pi} \int_0^\infty k_n^2 dk_n
 {\cal T}_{\ell_n}^{{\cal X}_n}(k_n) j_{L_n}(k_n r)  \right] 
B^{\rm eq}_{k_1 k_2 k_3} 
~, \label{eq:CMB_bis}
\end{eqnarray}
where 
\begin{eqnarray}
I^{s_1 s_2 s_3}_{l_1 l_2 l_3} 
&\equiv& \sqrt{\frac{(2 l_1 + 1)(2 l_2 + 1)(2 l_3 + 1)}{4 \pi}}
\left(
  \begin{array}{ccc}
  l_1 & l_2 & l_3 \\
   s_1 & s_2 & s_3 
  \end{array}
 \right)~.
\end{eqnarray}
Here, selection rules of the Wigner symbols allow $L_n$ to run over $|\ell_n - 2|  \leq L_n \leq \ell_n + 2$ under the restrictions: $L_1 + L_2 + L_3 = {\rm even}$ and $|L_1 - L_2| \leq L_3 \leq L_1 + L_2$. On the other hand, concerning $\ell_n$, we stress that there is no restriction except a triangle inequality $|\ell_1 - \ell_2| \leq \ell_3 \leq \ell_1 + \ell_2$. This implies that the CMB bispectra have nonzero values for both $\ell_1 + \ell_2 + \ell_3 = {\rm even}$ and $\ell_1 + \ell_2 + \ell_3 = {\rm odd}$. Physically, this is a consequence of an asymmetric spin dependence of the tensor bispectrum, i.e., $|B^{\lambda_1 \lambda_2 \lambda_3}| \neq |B^{-\lambda_1 -\lambda_2 -\lambda_3}|$, and directly connected to the absence of $x$ in eq.~(\ref{eq:cmb_bis_form}).

\begin{figure}[t]
  \begin{tabular}{cc}
    \begin{minipage}{0.5\hsize}
  \begin{center}
    \includegraphics[width=7.5cm,clip]{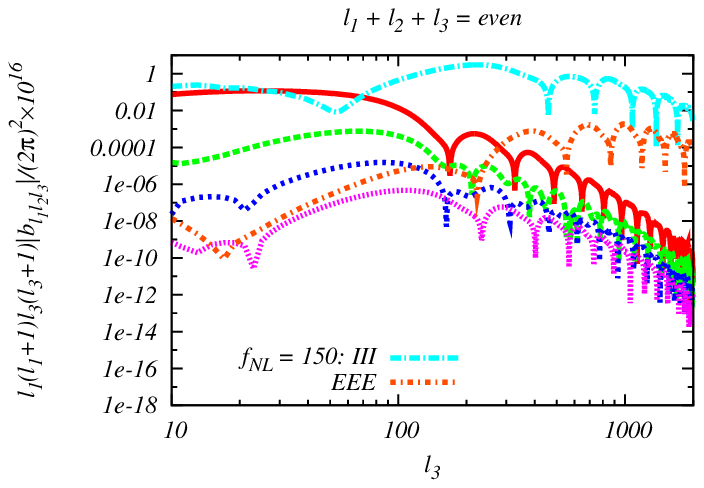}
  \end{center}
\end{minipage}
\begin{minipage}{0.5\hsize}
  \begin{center}
    \includegraphics[width=7.5cm,clip]{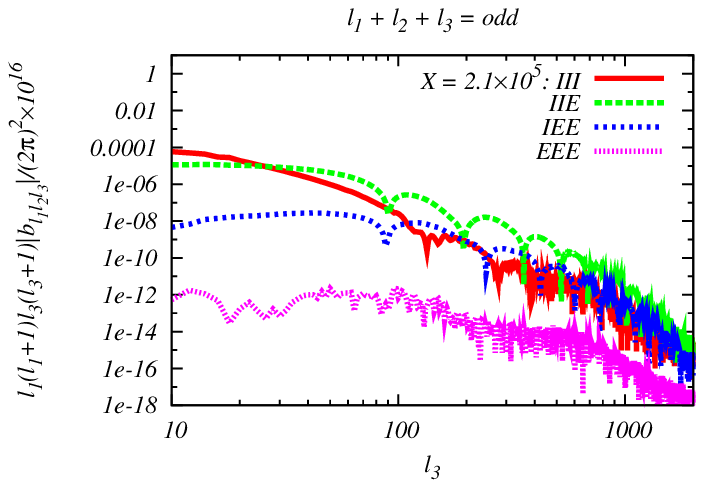}
  \end{center}
\end{minipage}
\end{tabular}
  \begin{tabular}{cc}
    \begin{minipage}{0.5\hsize}
  \begin{center}
    \includegraphics[width=7.5cm,clip]{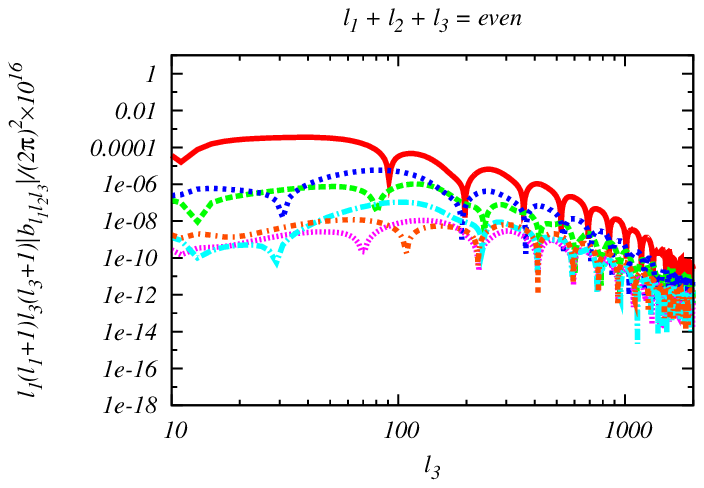}
  \end{center}
\end{minipage}
\begin{minipage}{0.5\hsize}
  \begin{center}
    \includegraphics[width=7.5cm,clip]{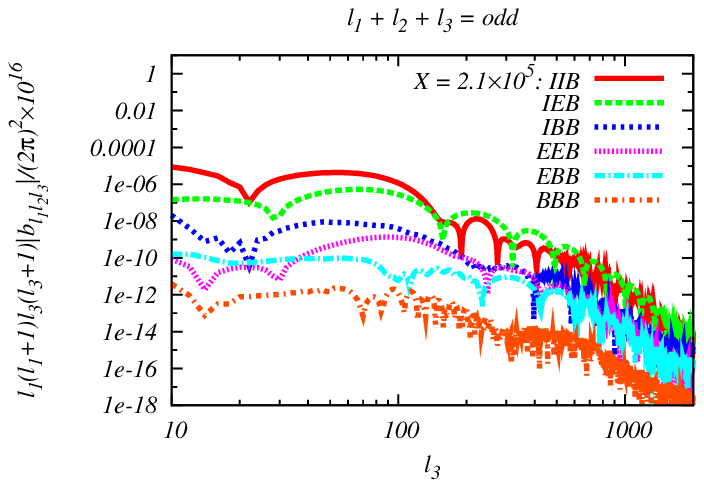}
  \end{center}
\end{minipage}
\end{tabular}
  \caption{All possible CMB bispectra, i.e., $\Braket{III}$, $\Braket{IIE}$, $\Braket{IEE}$ and $\Braket{EEE}$ (top two panels), and $\Braket{IIB}$, $\Braket{IEB}$, $\Braket{IBB}$, $\Braket{EEB}$, $\Braket{EBB}$ and $\Braket{BBB}$ (bottom two panels), induced by the tensor non-Gaussianity with $X = 2.1 \times 10^{5}$ and ${\cal P} = 2.5 \times 10^{-9}$ for $\ell_1 + 2 = \ell_2 + 1 = \ell_3$. Left and right two panels describe the parity-even ($\ell_1 + \ell_2 + \ell_3 = {\rm even}$) and parity-odd ($\ell_1 + \ell_2 + \ell_3 = {\rm odd}$) components, respectively. For comparison, we also plot $\Braket{III}$ and $\Braket{EEE}$ from the equilateral non-Gaussianity with $f_{\rm NL} = 150$. Other cosmological parameters are fixed using the {\it Planck} results \cite{Ade:2013lta}. The parity-odd bispectra seem to oscillate rapidly since they hate symmetric signals as $\ell_1 \sim \ell_2 \sim \ell_3$.} \label{fig:CMB_bis}
\end{figure}

Figure~\ref{fig:CMB_bis} depicts reduced bispectra given by eq.~(\ref{eq:CMB_bis}) of the temperature, E-mode and B-mode anisotropies for $\ell_1 \approx \ell_2 \approx \ell_3$, which is defined as 
\begin{eqnarray}
b_{\ell_1 \ell_2 \ell_3}^{{\cal X}_1 {\cal X}_2 {\cal X}_3} 
&=& G_{\ell_1 \ell_2 \ell_3}^{-1} B_{\ell_1 \ell_2 \ell_3}^{{\cal X}_1 {\cal X}_2 {\cal X}_3} ~, \\ 
G_{\ell_1 \ell_2 \ell_3} 
&\equiv& \frac{1}{6} \left[ \frac{2 \sqrt{\ell_3 (\ell_3 + 1) \ell_2 (\ell_2 +
1)}}{\ell_1(\ell_1 + 1) - \ell_2 (\ell_2 + 1) - \ell_3 (\ell_3 + 1)} 
\sqrt{\frac{\prod_{n=1}^3 (2 \ell_n + 1)}{4 \pi}}
\left(
  \begin{array}{ccc}
  \ell_1 & \ell_2 & \ell_3 \\
   0 & -1 & 1
  \end{array}
 \right) \right. \nonumber \\ 
&&\left. \qquad+ {5~\rm perms.} \right] ~.
\end{eqnarray} 
Note that $G_{\ell_1 \ell_2 \ell_3} = I_{\ell_1 \ell_2 \ell_3}^{0~0~0}$ holds if $\ell_1 + \ell_2 + \ell_3 = {\rm even}$. In figure~\ref{fig:CMB_bis}, the usual equilateral bispectra with $f_{\rm NL} = 150$ are also plotted, and it seems to be comparable in magnitude to the tensor bispectra with $X = 2.1 \times 10^{5}$ for $\ell \lesssim 100$. This relation has also been confirmed in the flat-sky analysis \cite{Cook:2013xea}. In the tensor bispectra, we can see the characteristic signatures associated with the tensor-mode CMB fields as mentioned above, i.e., the ISW enhancement in temperature for $\ell \lesssim 100$ and a peak due to Thomson scattering in the polarization at $\ell \sim 100$. Generally, in the tensor mode, the temperature fluctuations are larger than the polarization ones, and E and B modes have almost same amplitudes \cite{Pritchard:2004qp}. Such a magnitude relationship seems to hold in this figure too. With a confirmation of both parity-even ($\ell_1 + \ell_2 + \ell_3 = {\rm even}$) and parity-odd ($\ell_1 + \ell_2 + \ell_3 = {\rm odd}$) signals in all types of the CMB bispectra, These support the validity of our computations. 

While in the cross-bispectra the parity-odd signals are comparable in magnitude to the parity-even ones, the parity-odd auto-bispectra, i.e., $\Braket{III}$, $\Braket{EEE}$ and $\Braket{BBB}$, are slightly smaller than the parity-even counterparts. This is because of antisymmetry of the parity-odd CMB bispectrum, which means that three fields forming the bispectrum cannot take the identical states each other, e.g., $B_{\ell \ell \ell}^{III} = B_{\ell \ell \ell'}^{EEE} = 0$ \cite{Shiraishi:2011st, Shiraishi:2012sn}. This suppression may slightly decrease the total signals from the parity-odd bispectra in comparison with the parity-even case, as seen in the next section.

\section{Detectability analysis}\label{sec:Fisher}

To discuss detectability of the above bispectrum signals due to the tensor non-Gaussianity, in this section we evaluate error bars of $X$ (\ref{eq:X_def}) using the Fisher matrix. We are then interested in $X \lesssim {\cal O}(10^5)$. In this region, gauge-field-induced curvature perturbations are negligible and therefore ${\cal P}$ will coincide with observed power spectrum of curvature perturbations \cite{Cook:2013xea}. Accordingly, we here adopt ${\cal P} = 2.5 \times 10^{-9}$. As instrumental noise information, we adopt the data expected from the ${\it Planck}$ and the proposed PRISM experiments \cite{:2006uk, Andre:2013afa}. Computational methodology for the Fisher forecast is based on ref.~\cite{Shiraishi:2013vha}.

\subsection{Temperature and E-mode bispectra}

Let us start from an estimation for the temperature and E-mode bispectra. In this case, we shall analyze under a contamination of the usual equilateral non-Gaussianity since its CMB bispectra are also amplified at $\ell_1 \sim \ell_2 \sim \ell_3$. The contamination by the equilateral non-Gaussianity appears only in the parity-even space ($\ell_1 + \ell_2 + \ell_3 = {\rm even}$). Then, each element of the Fisher matrix can be defined as
\begin{eqnarray}
F_{ij} = \sum_{\substack{{\cal X}_1 {\cal X}_2 {\cal X}_3 \\ {\cal X}_1' {\cal X}_2' {\cal X}_3'}} \sum_{\ell_1 \leq \ell_2 \leq \ell_3 \leq \ell_{\rm max}} 
\frac{1}{\Delta_{\ell_1 \ell_2 \ell_3}} 
\tilde{B}_{i,  \ell_1 \ell_2 \ell_3}^{{\cal X}_1' {\cal X}_2' {\cal X}_3'}
\left[\prod_{n=1}^3 (C^{-1})_{\ell_n}^{{\cal X}_n {\cal X}_n'} \right]
\tilde{B}_{j, \ell_1 \ell_2 \ell_3}^{{\cal X}_1 {\cal X}_2 {\cal X}_3} ~,
\end{eqnarray}
where  
\begin{eqnarray}
\Delta_{\ell_1 \ell_2 \ell_3} = (-1)^{\ell_1 + \ell_2 + \ell_3} 
(1 + 2 \delta_{\ell_1, \ell_2} \delta_{\ell_2, \ell_3})  
+ \delta_{\ell_1, \ell_2} + \delta_{\ell_2, \ell_3} + \delta_{\ell_3, \ell_1}~, 
\end{eqnarray}
and ${\cal X}_1 {\cal X}_2 {\cal X}_3$ or ${\cal X}_1' {\cal X}_2' {\cal X}_3'$ runs over 8 combinations: $III$, $IIE$, $IEI$, $EII$, $IEE$, $EIE$, $EEI$ and $EEE$. Note that this formula is applicable to not only the parity-even space but also the parity-odd one ($\ell_1 + \ell_2 + \ell_3 = {\rm odd}$). The inverse matrix of the power spectrum reads 
\begin{eqnarray}
(C^{-1})_{\ell}^{\cal XX'} \equiv 
\left(
  \begin{array}{cc}
  C_{\ell}^{II} & C_{\ell}^{IE}  \\
  C_{\ell}^{EI} & C_{\ell}^{EE}
  \end{array}
 \right)^{-1} ~, 
\end{eqnarray}
where $C_\ell^{\cal XX'}$ is the sum of the cosmic variance spectrum and the noise spectrum \cite{Shiraishi:2013vha}. $\tilde{B}_i$ and $\tilde{B}_j$ consist of normalized CMB bispectra generated from the tensor non-Gaussianity ($\tilde{B}_{p} \equiv B_{p} / X^3$) and the equilateral curvature non-Gaussianity ($\tilde{B}_{e} \equiv B_{e} / f_{\rm NL} $). If we set a 2-dimensional Fisher matrix as 
 \begin{eqnarray}
{}^{(2)}F &=& \left(
  \begin{array}{cc}
  F_{pp} & F_{pe}  \\
  F_{ep} & F_{ee}
  \end{array}
 \right)~, \label{eq:Fisher2D}
\end{eqnarray} 
the $1\sigma$ error bars are expressed as 
\begin{eqnarray}
\left(\delta (X^3), \delta f_{\rm NL}\right) 
= \left( \sqrt{{}^{(2)}F^{-1}_{11}}, \sqrt{{}^{(2)}F^{-1}_{22}} \right) ~. \label{eq:X3_fnl}
\end{eqnarray}

\begin{figure}[t]
  \begin{tabular}{cc}
    \begin{minipage}{0.5\hsize}
  \begin{center}
    \includegraphics[width=7.5cm,clip]{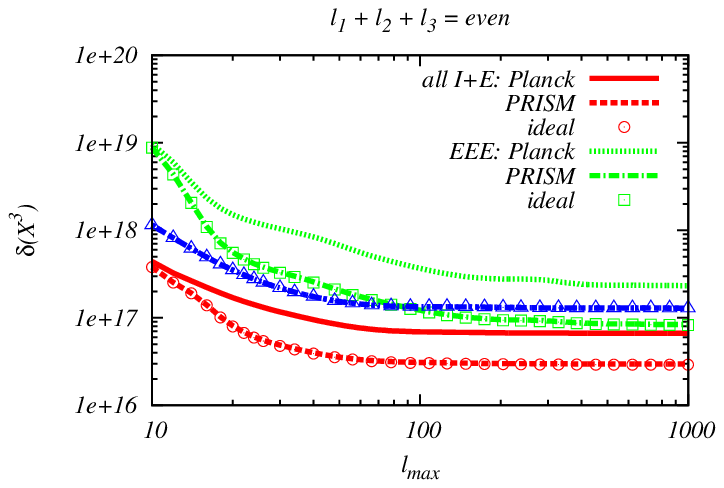}
  \end{center}
\end{minipage}
\begin{minipage}{0.5\hsize}
  \begin{center}
    \includegraphics[width=7.5cm,clip]{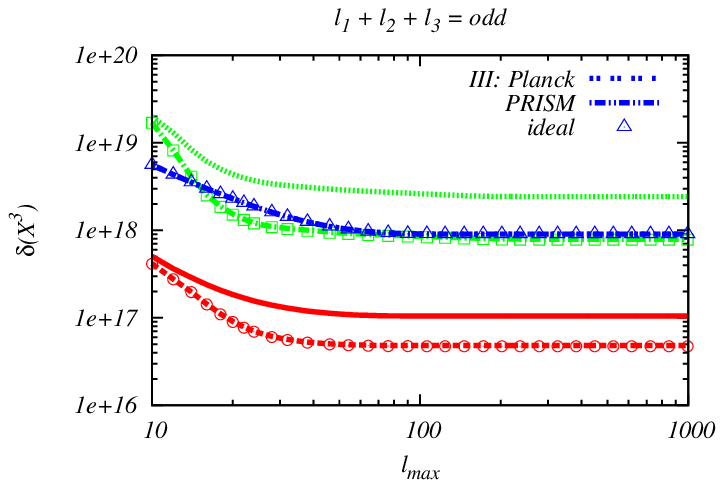}
  \end{center}
\end{minipage}
\end{tabular}
  \caption{Expected $1\sigma$ errors of $X^3$ (\ref{eq:X3_fnl}) obtained by using the parity-even (left panel) and parity-odd (right panel) signals in all types of the temperature and E-mode bispectra (red lines), the E-mode auto-bispectrum alone (green lines) and the temperature auto-bispectrum alone (blue lines). Here we assume the {\it Planck}, PRISM, and cosmic-variance-limited ideal experiments.} \label{fig:error_I+E}
\end{figure} 

Figure~\ref{fig:error_I+E} depicts $\delta (X^3)$ as functions of $\ell_{\rm max}$ estimated from all combinations of the temperature and E-mode bispectra, the E-mode auto-bispectrum alone and the temperature auto-bispectrum alone, respectively. Here, we display results from the parity-even and parity-odd spaces separately. From this figure, we can notice that $\delta (X^3)$ saturates for $\ell_{\rm max} \gtrsim 100$ in every case. This is due to rapid decays of the tensor temperature and polarization bispectra for $\ell \gtrsim 100$ (see figure~\ref{fig:CMB_bis}). Concerning features associated with parity, one can find that the error bars from the parity-odd signals are larger than those from the parity-even signals in the $\Braket{III}$ and $\Braket{EEE}$ cases. This is a consequence of the suppression of the auto-bispectra as mentioned in section~\ref{sec:CMB_bis}. Regardless of it, owing to contributions of the 6 cross-bispectra, the errors estimated from all possible 8 bispectra are comparable to or slightly smaller than the parity-even counterparts.

\begin{table}[t]
\begin{center}
  \begin{tabular}{|c||c|c|c|c|c|c|c|} \hline
     & $III$ & $EEE$ & all $I+E$ & $BBB$ ($r = 0.05$) & $BBB$ ($ r = 5 \times 10^{-4}$) \\ \hline 
    {\it Planck} & 127 (129) & 232 (233) & 56 (65) & 17 (19) & 2.1 (2.1)  \\ 
    PRISM & 127 (129) & 83 (84) & 25 (30) & 0.87 (1.0) & 0.015 (0.017) \\
    ideal & 127 (129) & 82 (83) & 25 (29) & 0.12 (0.20) & $1.2 ~ (2.0) \times 10^{-4}$ \\ \hline
  \end{tabular}
\end{center}
\caption{Expected $1\sigma$ errors of $X^3$ normalized by $10^{15}$ in the $III$, $EEE$, all $I+E$ cases ($\ell_{\rm max} = 1000$) and the $BBB$ case ($\ell_{\rm max} = 500$) for each experiment. The tensor-to-scalar ratio $r$ determines the amplitude of the B-mode cosmic variance spectrum. Here we summarize the results estimated from both the parity-even and parity-odd signals. In addition, for comparison, the errors from the parity-even signals alone are written in parentheses.}\label{tab:X3}
\end{table}

Practical values of $\delta (X^3)$ at $\ell_{\rm max} = 1000$ are summarized in table~\ref{tab:X3}. Interestingly, if using full set of the temperature and E-mode bispectra in both the parity-even and parity-odd spaces, $\delta (X^3)$ can be 80\% reduced in comparison with the $\Braket{III}$ analysis under the cosmic-variance-limited ideal experiment. This seems to be a common feature of the tensor non-Gaussianity \cite{Shiraishi:2013vha}. Such a level of improvement cannot be attained in the scalar non-Gaussianity case, where $\delta f_{\rm NL}$ is only 50\% reduction (see appendix~\ref{appen:fnl}). This result shows the powerful of the polarization bispectra. In this table, we can notice that the parity-odd information also improves the errors if we use all types of the temperature and E-mode bispectra. These improvements yield $\delta (X^3) = 5.6 \times 10^{16}$ ({\it Planck}) and $2.5 \times 10^{16}$ (PRISM or ideal).

Finally, we shall mention the contamination of the usual equilateral non-Gaussianity. The correlation coefficient between the tensor temperature and E-mode bispectra in both the parity-even and parity-odd spaces and the parity-even equilateral ones reads
\begin{eqnarray}
\frac{F_{pe}}{\sqrt{F_{pp} F_{ee}}} = 0.036 ~,
\end{eqnarray}
which shows the lack of correlation, and hence they do not bias each other's error estimation. This is a consequence of the shape difference of the CMB fields between the scalar and tensor modes.

\subsection{B-mode bispectra}

Here, let us consider error estimations including the B-mode information. Such bispectra correspond to 6 additional contributions: $\Braket{IIB}$, $\Braket{IEB}$, $\Braket{IBB}$, $\Braket{EEB}$, $\Braket{EBB}$ and $\Braket{BBB}$ and considering all these information will improve $\delta (X^3)$ drastically. However, this is a very complicated procedure and hence here we focus only on the $\Braket{BBB}$ analysis. In this case, there is no contamination from the usual equilateral non-Gaussianity because of the absence of B-mode creation by the scalar mode. Therefore, we can estimate the error through 1-dimensional Fisher matrix as 
\begin{eqnarray}
F &\equiv& \sum_{\ell_1 \leq \ell_2 \leq \ell_3 \leq {\ell_{\rm max}}} 
\frac{ \left(\tilde{B}^{BBB}_{\ell_1 \ell_2 \ell_3}\right)^2}{\Delta_{\ell_1 \ell_2 \ell_3} 
\prod_{n=1}^3 C^{BB}_{\ell_n}} ~, \\
\delta (X^3) &=& F^{-1/2} ~,  \label{eq:X31D}
\end{eqnarray}
where $\tilde{B} \equiv B_p / X^3$. Here we take $\ell_{\rm max} \leq 500$ since lensing B-mode contribution may behave as a bias on small scales \cite{Lewis:2011fk, Hanson:2013hsb}. 

\begin{figure}[t]
  \begin{tabular}{cc}
    \begin{minipage}{0.5\hsize}
  \begin{center}
    \includegraphics[width=7.5cm,clip]{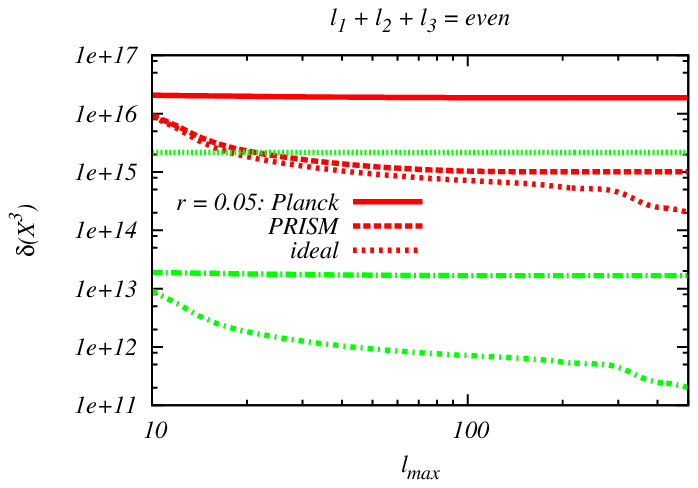}
  \end{center}
\end{minipage}
\begin{minipage}{0.5\hsize}
  \begin{center}
    \includegraphics[width=7.5cm,clip]{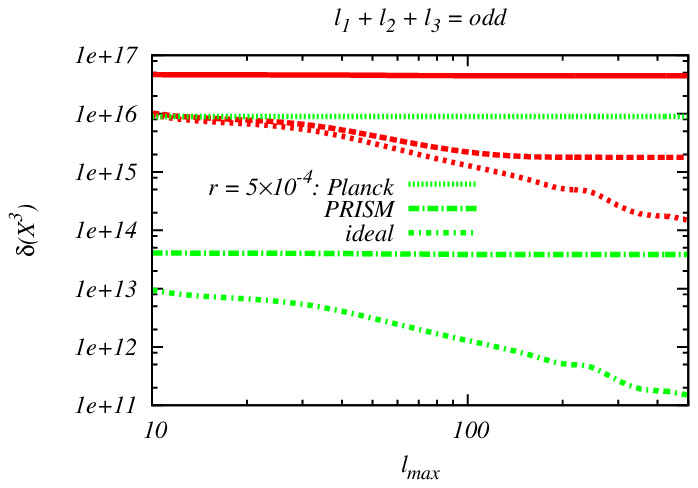}
  \end{center}
\end{minipage}
\end{tabular}
  \caption{Expected $1\sigma$ errors of $X^3$ (\ref{eq:X31D}) estimated by using the parity-even (left panel) and parity-odd (right panel) signals in the B-mode auto-bispectrum. As the cosmic variance spectra, we adopt the B-mode power spectra with $r = 0.05$ (red lines) and $5 \times 10^{-4}$ (green lines). Here we assume the {\it Planck}, PRISM, and cosmic-variance-limited ideal experiments.} \label{fig:error_BBB}
\end{figure} 
 
Numerical results of $\delta(X^3)$ are described in figure~\ref{fig:error_BBB}. We also summarize the values at $\ell_{\rm max} = 500$ in table~\ref{tab:X3}. As the cosmic variance spectra, we adopt the standard B-mode power spectra where corresponding tensor-to-scalar ratios are $r = 0.05$ and $5 \times 10^{-4}$. In the cosmic-variance-limited ideal experiment, the variance of the bispectrum is determined by $r$ alone and hence we can find a simple relationship $\delta(X^3) =  1.1 \times 10^{16} r^{3/2}$ at $\ell_{\rm max} = 500$ using both the parity-even and parity-odd signals. $\Braket{BBB}$ can improve the error more than $\Braket{EEE}$ despite the same noise spectrum (see appendix in ref.~\cite{Shiraishi:2013vha}) because of smallness of the B-mode cosmic variance spectrum. While the improvements of the parity-odd signals are slightly weaker than the parity-even signals for low $\ell_{\rm max}$, interestingly, this situation seems to be reversed for high $\ell_{\rm max}$ if the instrumental noise is negligible. Owing to these signatures, $\delta (X^3)$ can reach the values comparable to or less than $10^{15}$ in the {\it Planck} or the proposed PRISM experiment.

\section{Summary and discussion}\label{sec:summary}

A rolling pseudoscalar can induce large equilateral-type non-Gaussianity in the tensor perturbations via the dynamics of a chiral gauge field. Such non-Gaussianity violates parity invariance and imprints characteristic signatures in the CMB temperature and polarization bispectra.

In general, the parity-violating signatures in the non-Gaussianity appear in the complicated spin and angle dependence of resulting CMB bispectra. We dealt with these in the full-sky formalism and confirmed that all types of CMB temperature, E-mode and B-mode bispectra have nonzero signals for both $\ell_1 + \ell_2 + \ell_3 = {\rm even}$ and $\ell_1 + \ell_2 + \ell_3 = {\rm odd}$. This property cannot be seen in the flat-sky analysis because of lack of the discreteness in $\ell$ space. Physically, this reflects the asymmetric spin dependence of the tensor non-Gaussianity, namely $|B^{\lambda_1 \lambda_2 \lambda_3}| \neq |B^{-\lambda_1 -\lambda_2 -\lambda_3}|$. Numerical evaluations show that such CMB bispectra are amplified on large scales and take quite different shapes from the usual scalar equilateral bispectra due to the tensor-mode transfer functions. These mean that the existence of the usual equilateral bispectra does not bias the estimation of pseudoscalar signals.
 
We evaluated the detectability of these signals through the Fisher matrix analysis. Then, it was clarified that considering both the parity-even and parity-odd contributions improve the detectability more. The analysis of the temperature auto-bispectrum alone will detect a model parameter, defined in eq.~(\ref{eq:X_def}), $X = 5.0 \times 10^5$ at 68\% CL under the {\it Planck} or proposed PRISM experiment. Inclusion of the E-mode contributions and whole $\ell$-space analysis potentially improve of 400\% the detectability of the primordial tensor bispectrum ($\propto X^3$) with respect to the above temperature alone case. The corresponding $1\sigma$ values of $X$ read $3.8 \times 10^5$ and $2.9 \times 10^5$ under the {\it Planck} and PRISM experiments, respectively. Moreover, we presented the power of the B-mode bispectra to reduce the error drastically. If the instrumental noise is negligible and the B-mode cosmic variance can be expressed by the tensor-to-scalar ratio $r$, we can write the $1\sigma$ error of $X$ as $\delta X = 2.2 \times 10^5 \sqrt{r}$. In this sense, we will be able to observe $X$ less than $10^4$ if $r < 0.002$. In practice, the instrumental noises prevent $\delta X$ from becoming smaller, and under $r = 0.05$ $(5 \times 10^{-4})$, $X = 2.6~ (1.3) \times 10^5$ and $9.5~ (2.5) \times 10^4$ will be possible to be detected in the {\it Planck} and PRISM experiments, respectively. The CMB bispectra seem to have the detectability of $X$ comparable to or greater than the TB and EB correlations as shown in appendix~\ref{appen:TB_EB}.

In this paper, we focused on the CMB signatures originating from a rolling pseudoscalar. However, there also exist other sources which create the parity-violating tensor non-Gaussianities and imprint similar signatures in the CMB bispectra \cite{Shiraishi:2011st, Shiraishi:2012sn}. To differentiate between these sources, a more comprehensive analysis considering each contamination will be required. Then, the correlations $\Braket{IIB}$, $\Braket{IEB}$, $\Braket{IBB}$, $\Braket{EEB}$, $\Braket{EBB}$ and $\Braket{BBB}$ may also be informative. These are expected to be done in future papers.

\acknowledgments
We thank Sabino Matarrese and Marco Peloso for useful comments and discussions. MS was supported in part by a Grant-in-Aid for JSPS Research under Grant No.~25-573 and the ASI/INAF Agreement I/072/09/0 for the Planck LFI Activity of Phase E2. We also acknowledge the Kobayashi-Maskawa Institute for the Origin of Particles and the Universe, Nagoya University for providing computing resources that were useful in conducting the research reported in this paper.

\appendix
\section{Polarization vector and tensor} \label{appen:polarization}

In this paper, we utilize a divergenceless polarization vector $\epsilon_a^{(\pm 1)}$ and a transverse and traceless polarization tensor $e_{ab}^{(\pm 2)}$ satisfying \cite{Shiraishi:2010kd}
\begin{eqnarray}
\begin{split}
\hat{k}^a \epsilon_a^{(\lambda)}(\hat{\bf k}) &= 0~, \\ 
\eta^{abc} \hat{k}_a \epsilon_b^{(\lambda)}(\hat{\bf k}) 
&= -\lambda i \epsilon_c^{(\lambda)}(\hat{\bf k}) ~, \\ 
\epsilon^{(\lambda) *}_a (\hat{\bf k}) &= \epsilon^{(-\lambda)}_a (\hat{\bf k})
 = \epsilon^{(\lambda)}_a (-\hat{\bf k})~, \\
\epsilon^{(\lambda)}_a (\hat{\bf k}) \epsilon^{(\lambda')}_a (\hat{\bf k}) 
&= \delta_{\lambda, -\lambda'}  ~, 
\end{split}
\end{eqnarray}
and 
\begin{eqnarray}
\begin{split}
e_{ab}^{(\lambda)}(\hat{\bf k}) &\equiv \sqrt{2} \epsilon_a^{(\frac{\lambda}{2})}(\hat{\bf k}) \epsilon_b^{(\frac{\lambda}{2})}(\hat{\bf k}) ~, \\ 
e_{aa}^{(\lambda)}(\hat{\bf k}) &= \hat{k}_a e_{ab}^{(\lambda)}(\hat{\bf k}) = 0~, \\
e_{ab}^{(\lambda) *}(\hat{\bf k}) &= e_{ab}^{(-\lambda)}(\hat{\bf k}) = e_{ab}^{(\lambda)}(- \hat{\bf k})~, \\
e_{ab}^{(\lambda)}(\hat{\bf k}) e_{ab}^{(\lambda')}(\hat{\bf k}) &= 2
\delta_{\lambda, -\lambda'} ~, \label{eq:pol_tens_relation} 
\end{split}
\end{eqnarray}
where $\eta^{abc}$ is a 3-dimensional antisymmetric tensor normalized by $\eta^{123} = 1$. An expression in $\ell$ space is convenient, reading 
\begin{eqnarray}
e_{ab}^{(\lambda)} (\hat{\bf k}) 
= \frac{3}{\sqrt{2 \pi}}  
\sum_{M m_a m_b} {}_{-\lambda}Y_{2 M}^*(\hat{\bf k}) 
\alpha^{m_a}_{a} \alpha^{m_b}_b 
\left(
  \begin{array}{ccc}
  2 & 1 &  1\\
  M & m_a & m_b 
  \end{array}
\right) ~,
\end{eqnarray}
with
\begin{eqnarray}
\alpha_a^m \alpha_a^{m'} = \frac{4 \pi}{3} (-1)^m \delta_{m,-m'}~, \ \
\alpha_a^m \alpha_a^{m' *} = \frac{4 \pi}{3} \delta_{m,m'}~.
\end{eqnarray}
Then, dealing with the Wigner symbols yields  
\begin{eqnarray}
e_{ab}^{(-\lambda_1)}(\hat{\bf k_1}) e_{bc}^{(-\lambda_2)}(\hat{\bf k_2}) e_{ca}^{(-\lambda_3)}(\hat{\bf k_3})
= -\frac{( 8\pi )^{3/2} }{10} \sqrt{\frac{7}{3}}
\left[ \prod_{n=1}^3 \sum_{\mu_n} {}_{\lambda_n}Y_{2 \mu_n}^*(\hat{\bf k_n})  \right]
\left(
  \begin{array}{ccc}
   2 & 2 & 2 \\
  \mu_1 & \mu_2 & \mu_3
  \end{array}
 \right).
\end{eqnarray}
With a multipole expansion of the delta function:
\begin{eqnarray}
\delta^{(3)}\left( \sum_{n=1}^3 {\bf k_n} \right) 
&=& 8 \int_0^\infty r^2 dr 
\left[ \prod_{n=1}^3 \sum_{L_n M_n} 
 (-1)^{\frac{L_n}{2}} j_{L_n}(k_n r) 
Y_{L_n M_n}^*(\hat{\bf k_n}) \right] 
\nonumber \\
&&\times 
I_{L_1 L_2 L_3}^{0 \ 0 \ 0}
 \left(
  \begin{array}{ccc}
  L_1 & L_2 & L_3 \\
  M_1 & M_2 & M_3 
  \end{array}
 \right)~,
\end{eqnarray}
this representation is applied to formulation of the CMB bispectrum in section~\ref{sec:CMB_bis}.

\section{Errors of the equilateral non-Gaussianity}\label{appen:fnl}

\begin{figure}[t]
  \begin{center}
    \includegraphics[width=12cm,clip]{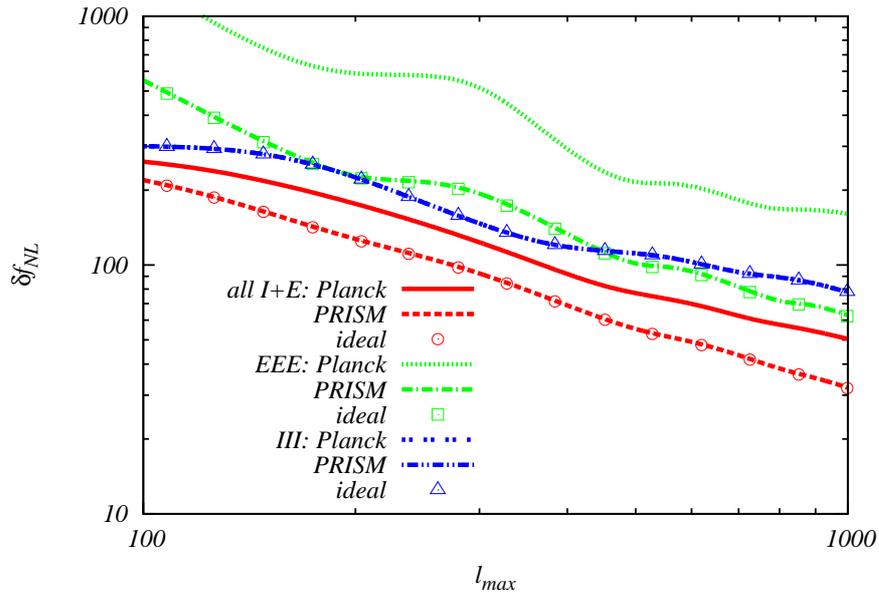}
  \end{center}
  \caption{Expected $1\sigma$ errors of $f_{\rm NL}$ (\ref{eq:X3_fnl}) estimated from all information of the temperature and E-mode bispectra (red lines), $\Braket{EEE}$ (green lines) and $\Braket{III}$ (blue lines) in the {\it Planck}, PRISM and ideal experiments.} \label{fig:error_fnl}
\end{figure}

In figure~\ref{fig:error_fnl}, we plot $1\sigma$ errors of the equilateral nonlinearity parameter in the 2-dimensional Fisher matrix analysis with the parity-violating tensor non-Gaussianity. It is observed that the analysis with all types of the temperature and E-mode bispectra leads to 50\% reduction of $\delta f_{\rm NL}$ in comparison with the analysis of $\Braket{III}$ alone under the PRISM or ideal experiment. Thanks to uncorrelation with the tensor non-Gaussianity, $\delta f_{\rm NL}$ is in good agreement with $F_{ee}^{-1/2}$ of eq.~(\ref{eq:Fisher2D}).

\section{TB and EB correlations}\label{appen:TB_EB}

\begin{figure}[t]
    \begin{center}
    \includegraphics[width=12cm,clip]{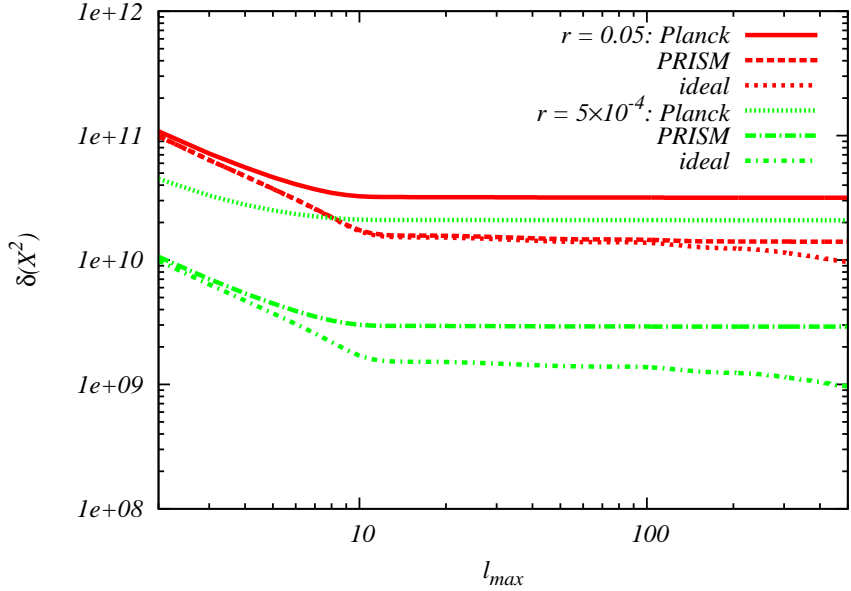}
  \end{center}
  \caption{Expected $1\sigma$ errors of $X^2$ estimated from the TB and EB correlations. Here we adopt the cosmic variance spectra and the noise spectra used in figure~\ref{fig:error_BBB}.} \label{fig:error_IB+EB}
\end{figure} 

Here, we shall discuss the detectability of the parameter $X$ from TB and EB power spectra (hereinafter noted as $C_\ell^{IB}$ and $C_\ell^{EB}$, respectively). The TB or EB correlation is sourced by the difference between $\lambda = +2$ and $-2$ tensor power spectra, given as $\Braket{h_{\bf k_1}^{(+2)}h_{\bf k_2}^{(+2)}} - \Braket{h_{\bf k_1}^{(-2)}h_{\bf k_2}^{(-2)}} \equiv \Delta_h(k_1) \delta^{(2)}({\bf k_1} + {\bf k_2})$, reading
\begin{eqnarray}
C_{\ell}^{IB/EB}
&=& \frac{2}{\pi} \int_0^\infty k^2 dk 
  {\cal T}_{\ell}^{I/E}(k) 
  {\cal T}_{\ell}^{B}(k) 
\Delta_h(k)~.
\end{eqnarray}
In our case, gravitational waves are positively-polarized and hence $\Delta_h$ is dominated by the $+2$ power spectrum: \cite{Sorbo:2011rz} 
\begin{eqnarray}
\Delta_h(k) \approx 8.6 \times 10^{-7} \frac{H^4}{M_P^4} \frac{e^{4\pi\xi}}{\xi^6} k^{-3} \propto X^2  ~.
\end{eqnarray}
Concrete shapes of $C_\ell^{IB}$ and $C_\ell^{EB}$ can be seen in ref.~\cite{Gluscevic:2010vv}. 

Let us define a Fisher matrix for an error estimation of $X^2$ as \cite{Gluscevic:2010vv}
\begin{eqnarray}
F = \sum_{\ell = 2}^{\ell_{\rm max}} \sum_{ij} 
\frac{\partial C_{\ell}^{i}}{\partial (X^2)} 
(Cov^{-1})_\ell^{ij} \frac{\partial C_{\ell}^{j}}{\partial (X^2)} ~,
\end{eqnarray}
where $i$ or $j$ runs over $IB$ and $EB$, and a 2-dimensional covariance matrix is given by
\begin{eqnarray}
Cov_\ell^{ij} &=& \frac{1}{2\ell +1 }
\left(
\begin{array}{ccc}
  C_\ell^{II} C_\ell^{BB} & C_\ell^{IE} C_\ell^{BB} \\
  C_\ell^{IE} C_\ell^{BB} & C_\ell^{EE} C_\ell^{BB}
\end{array}
\right) ~.
\end{eqnarray}
Here, we have obeyed a null hypothesis of the cosmic variance and instrumental noise from the TB and EB correlations. Figure~\ref{fig:error_IB+EB} describes expected $1 \sigma$ errors of $X^2$, which is calculated by $\delta (X^2) = F^{-1/2}$, under the presence of the B-mode cosmic variance spectra with $r = 0.05$ and $5 \times 10^{-4}$. These results are compatible with the previous works \cite{Gluscevic:2010vv}. We can also observe the rapid reduction of $\delta (X^2)$ for $\ell < 10$ thanks to large-scale information of the TB power spectrum, as discussed in ref.~\cite{Gluscevic:2010vv}.

This figure indicates that $X = 1.8 ~ (1.4) \times 10^5$ or $1.2 ~(0.54) \times 10^5$ will be detected at 68\% CL in the {\it Planck} or PRISM experiment when $r = 0.05$ $(5 \times 10^{-4})$. These values are comparable to or somewhat smaller than the results from the temperature and E-mode bispectra, and slightly larger than those from the B-mode bispectrum analysis (see section~\ref{sec:summary}). In the cosmic-variance-limited experiment, the $1\sigma$ error bar of $X$ is determined by $r^{1/4}$, reading $\delta X \approx 2.1 \times 10^5 r^{1/4}$. Because of the difference of $r$ dependence, the detectability of $X$ may be weaker than the B-mode bispectrum case for smaller $r$ as favored by current or future observations. 

\bibliography{paper}
\end{document}